\documentclass[conference]{IEEEtran}
\IEEEoverridecommandlockouts
\usepackage{cite}
\usepackage{amsmath,amssymb,amsfonts,amsthm}
\usepackage{algorithmic}
\usepackage{graphicx}
\usepackage{textcomp}
\usepackage[]{authblk}
\usepackage{xcolor}
\usepackage{comment}
\def\BibTeX{{\rm B\kern-.05em{\sc i\kern-.025em b}\kern-.08em
    T\kern-.1667em\lower.7ex\hbox{E}\kern-.125emX}}
    
\usepackage{graphicx}	
\usepackage{amsmath}	
\usepackage{comment}
\usepackage{xspace} 
\usepackage{mathtools}
\usepackage{stmaryrd}
\usepackage{hyperref}
\usepackage{float}
\usepackage{bm}
\usepackage{pdflscape}
\usepackage{pdfpages}
\usepackage[font=small,skip=4pt]{caption}
\DeclareFontFamily{U}{mathx}{}
\DeclareFontShape{U}{mathx}{m}{n}{<-> mathx10}{}
\DeclareSymbolFont{mathx}{U}{mathx}{m}{n}
\DeclareMathAccent{\widecheck}{0}{mathx}{"71}

\usepackage{multirow} \usepackage{threeparttable}
\usepackage{booktabs} \usepackage{stfloats}

\newcommand*{\V}[1]{\boldsymbol{#1}}   
\newcommand*{\M}[1]{\mathbf{#1}}       
\newcommand*{\TransposeLetter}{\hspace*{-.25ex}\top\hspace*{-.25ex}}
\newcommand*{\T}{^{\TransposeLetter}} 
\DeclareFontFamily{U}{mathx}{\hyphenchar\font45}
\DeclareFontShape{U}{mathx}{m}{n}{<-> mathx10}{}
\DeclareSymbolFont{mathx}{U}{mathx}{m}{n}

\DeclarePairedDelimiterX{\Paren}[1]{(}{)}{#1}
\DeclarePairedDelimiterX{\Brace}[1]{\{}{\}}{#1}
\DeclarePairedDelimiterX{\Brack}[1]{[}{]}{#1}
\DeclarePairedDelimiterX{\Abs}[1]{\rvert}{\lvert}{#1}
\DeclarePairedDelimiterX{\Norm}[1]{\lVert}{\rVert}{#1}
\DeclarePairedDelimiterX{\Avg}[1]{\langle}{\rangle}{#1}
\DeclarePairedDelimiterX{\Round}[1]{\lfloor}{\rceil}{#1}
\DeclarePairedDelimiterX{\Floor}[1]{\lfloor}{\rfloor}{#1}
\DeclarePairedDelimiterX{\Ceil}[1]{\lceil}{\rceil}{#1}
\DeclarePairedDelimiterX{\Inner}[2]{\langle}{\rangle}{#1,#2}

\DeclareMathOperator*{\argmin}{argmin}

\DeclareMathOperator{\Trace}{tr}

\DeclarePairedDelimiterXPP{\Expect}[1]{\mathbb{E}}(){}{#1}

\newcommand{\Sest}{\widehat{\M{S}}} 
\newcommand{\Fest}{\widehat{\M{F}}}
\newcommand{\Sigest}{\widehat{\M{\Sigma}}}

\usepackage[squaren,Gray]{SIunits}

\usepackage{numprint}

\makeatletter
\def\widebreve{\mathpalette\wide@breve}
\def\wide@breve#1#2{\sbox\z@{$#1#2$}%
     \mathop{\vbox{\m@th\ialign{##\crcr
\kern0.08em\brevefill#1{0.8\wd\z@}\crcr\noalign{\nointerlineskip}%
                    $\hss#1#2\hss$\crcr}}}\limits}
\def\brevefill#1#2{$\m@th\sbox\tw@{$#1($}%
  \hss\resizebox{#2}{\wd\tw@}{\rotatebox[origin=c]{90}{\upshape(}}\hss$}
\makeatletter

\begin{document}

\title{Shrinkage MMSE estimators of covariances beyond the zero-mean and stationary variance assumptions}

\author{Olivier Flasseur$^1$}
\author{{\'E}ric Thiébaut$^1$}
\author{Loïc Denis$^2$}
\author{Maud Langlois$^1$}
\affil{\small $^1$Centre de Recherche Astrophysique de Lyon, CNRS, Univ. Lyon, Univ. Claude Bernard Lyon 1, ENS Lyon, France \protect\\ $^2$Univ. Lyon, UJM-Saint-Etienne, CNRS, Institut d Optique Graduate School, Laboratoire Hubert Curien, France}

\maketitle

\begin{abstract}
	We tackle covariance estimation in low-sample scenarios, employing a structured covariance matrix with shrinkage methods. These involve convexly combining a low-bias/high-variance empirical estimate with a biased regularization estimator, striking a bias-variance trade-off. Literature provides optimal settings of the regularization amount through risk minimization between the true covariance and its shrunk counterpart. Such estimators were derived for zero-mean statistics with i.i.d. diagonal regularization matrices accounting for the average sample variance solely. We extend these results to regularization matrices  accounting for the sample variances both for centered and non-centered samples. In the latter case, the empirical estimate of the true mean is incorporated into our shrinkage estimators. Introducing confidence weights into the statistics also enhance estimator robustness against outliers. We compare our estimators to other shrinkage methods both on numerical simulations and on real data to solve a detection problem in astronomy.  
\end{abstract}

\begin{IEEEkeywords}
covariance estimation, shrinkage, regularization, bias-variance trade-off, inverse problems, detection
\end{IEEEkeywords}

\section{Introduction}
\label{sec:introduction}

\noindent Covariance estimation is crucial in diverse applications such as portfolio optimization \cite{ledoit2004well}, gene expression analysis \cite{schafer2005shrinkage}, and radar imaging  \cite{vasile2009coherency}. The latter is a typical example where inverse-problems are leveraged to solve detection or reconstruction tasks, considering measurement statistics through covariances \cite{maronna2019robust}. Limited data in high dimensions introduces a bias-variance trade-off in covariance estimation. Regularization of the noisy empirical covariance, often by imposing a specific structure, becomes necessary. \textit{Diagonal loading} involves adding a fraction of the identity matrix to reduce estimation variance (at the expense of increased bias). \textit{Shrinkage} methods generalize this, replacing the identity regularization matrix with a low-variance measure of data variability. The regularization amount crucially impacts the bias-variance trade-off and can be optimally estimated in an unsupervised manner by minimizing risk between the true unknown covariance and its shrunk estimate. 
Closed-form expressions are reported for several distributions \cite{chen2010shrinkage, ollila2019optimal}, but all involve shrinkage accounting solely for average sample variance.

\textit{Our contributions:} In the multi-variate Gaussian case, we propose to extend these shrinkage methods to include a regularization matrix accounting for the sample variances (i.e. with non-uniform diagonal values) both for centered and non-centered samples. In the latter case, the empirical estimate of the true unknown mean is incorporated into the derived closed-form expressions of the shrinkage.

Sect. \ref{sec:problem_statement} formulates the estimation problem. Our proposed estimators of the shrinkage coefficients are derived in Sects. \ref{sec:shrinkage_non_iid}-\ref{sec:sample_mean}. They are compared to state-of-the-art shrinkage estimators in Sect. \ref{sec:results} through numerical simulations. We finally apply them on real data to solve a detection problem in astronomy.

\section{Problem statement}
\label{sec:problem_statement}

\noindent We consider the general case of $N$ mutually independent and identically distributed $P$-dimensional real-valued Gaussian samples $\V x_n \in \mathbb{R}^P \thicksim \mathcal{N}(\V \mu, \M C)$, for $n \in \llbracket 1; N \rrbracket$. We define the weighted sample mean $\widehat{\V \mu}$ and the weighted sample covariance $\widehat{\M S}$ as respective estimators of $\V \mu$ and $\M C$:
\begin{equation}
	\widehat{\V \mu} = \frac{\sum_{n} \alpha_n \, \V x_n}{\sum_{n} \alpha_n} ~~~~\text{and}~~~~ \widehat{\M S} = \frac{\sum_{n} \beta_n (\V x_n - \widehat{\V \mu}) (\V x_n - \widehat{\V \mu})\T }{\sum_{n} \beta_n}\,,
	\label{eq:sample_mean_covariances}
\end{equation}
with $\lbrace \alpha_n \rbrace_{n=1:N}$ and $\lbrace \beta_n \rbrace_{n=1:N}$ two collections of fixed non-negative weights. The expectation of the estimators (\ref{eq:sample_mean_covariances}) are:
\begin{equation}
	\mathbb{E}(\widehat{\V \mu}) = \V \mu ~~~~\text{and}~~~~ \mathbb{E}(\widehat{\M S}) = (1 - \epsilon)\, \M C\,,
	\label{eq:exceptation_sample_mean_covariances}
\end{equation}
with $\epsilon \in \left[ 0, 1 \right]$ given by:
\begin{equation}
	\epsilon = \frac{2 \sum_{n} \alpha_n \beta_n}{\left( \sum_{n} \alpha_n
   \right)  \left( \sum_{n} \beta_n \right)} - \frac{\sum_{n} \alpha_n^2}{\left(
   \sum_{n} \alpha_n \right)^2}\,.
   \label{eq:epsilon}
\end{equation}
As $\widehat{\M S}$ is a biased estimator of $\M C$, we introduce the bias compensating factor $\gamma = (1-\epsilon)^{-1}$ when needed (note that $\alpha_n = \beta_n = 1\,, \forall n$ leads to the classical correction $\gamma= N / (N-1)$). 
In this context, we address the challenge of estimating the covariance matrix $\M C$ from a limited number $N$ of samples. The empirical sample estimator $\widehat{\M S}$ becomes highly noisy (when $N \simeq P$) and even rank-deficient (in particular when $N < P$). This introduces significant variance to the estimate $\widehat{\M S}$, which is detrimental in any inverse problem involving covariance matrix inversion, as discussed in Sect. \ref{sec:introduction}. Irrespective of the task, regularization of $\Sest$ is thus essential. One approach is to enhance the sample covariance matrix through shrinkage, where the resulting estimator $\widehat{\M C}$ is a convex combination of the low-bias/high-variance estimator $\widehat{\M S}$ and of a high-bias/low-variance estimator $\widehat{\M F}$ (possessing fewer degrees of freedom compared to $\widehat{\M S}$):
\begin{equation}
	\widehat{\M C} =  \gamma ((1 - \rho) \widehat{\M S} + \rho \widehat{\M F})\,,
	\label{eq:shrinkage_covariance}
\end{equation}
with $\rho \in \left[ 0, 1 \right]$ a coefficient setting the bias-variance trade-off. 
This combination of estimators with complementary properties originated in seminal works \cite{ledoit2004well}. An optimal $\rho$ can be defined through risk minimization  \cite{chen2010shrinkage}:
\begin{equation}
	\rho^{\text{OS}} = \argmin_{\rho \in \left[ 0, 1 \right]} \, \mathbb{E}(\lVert \widehat{\M C} - \M C \rVert_\text{F}^2) = \frac{\mathbb{E}(\Trace ( (\M C - \gamma\widehat{\M S})(\widehat{\M F} - \widehat{\M S}) )}{\gamma\mathbb{E}(\Trace  ((\widehat{\M F}-\widehat{\M S})^2 ))}\,,
	\label{eq:risk_minimization}
\end{equation}
where $\lVert . \rvert_\text{F}$ denotes the Frobenius norm.
This \textit{oracle shrinkage} (OS) estimator, dependent on the true (unknown) covariance $\M C$, cannot be used in practice. The work \cite{chen2010shrinkage} introduced a data-driven strategy to derive a practical \textit{oracle-approximating shrinkage} (OAS). This involves iteratively plugging a previous estimate $\widehat{\M C}$ to refine the resulting OAS estimate $\widehat{\rho}^{\text{ OAS}}$. Assuming centered samples ($\V \mu = \M 0$, thus $\widehat{\V \mu} = \M 0$, $\gamma = 1$ and $\alpha_n = 0, \forall n$), non-weighted sample covariances ($\beta_n = 1, \forall n$), 
and a diagonal structure for $\M F$ accounting solely for the mean variance of the samples (i.e., $\widehat{\M F} = (\Trace ( \widehat{\M S} )/P) \M I_P$) lead to the following estimate \cite{chen2010shrinkage}:
\begin{equation}
	\widehat{\rho}^{\text{ OAS}}_{1} = \frac{(1 - 2/P) \Trace ( \widehat{\M S}^2 ) + \Trace^2 ( \widehat{\M S} ) }{ (N+1-2/P) (\Trace ( \widehat{\M S}^2 ) - \Trace^2 (\widehat{\M S})/P  )  }\,.
	\label{eq:rho_chen}
\end{equation}
In the following, we extend these results by gradually relaxing the assumptions made in \cite{chen2010shrinkage} and by considering a finer estimator $\widehat{\M F}$ that better accounts for non-stationary variances: $[\widehat{\M F}]_{i j} = 0$ if $i \neq j$ and $[\widehat{\M F}]_{i i} = [\widehat{\M S}]_{i i}$ otherwise.

\section{Shrinkage for non-stationary variances}
\label{sec:shrinkage_non_iid}



\noindent We derive the expression of the OAS coefficient $\rho_2^{\text{OAS}}$ for a regularization matrix $\widehat{\M F}$ that accounts for the sample variances (instead of only the mean variance in \cite{chen2010shrinkage}). We still assume $\V \mu = \M 0$ (we set $\widehat{\V \mu} = \M 0$ in $\Sest$, and thus $\gamma=1$), and $\beta_n = 1, \forall n$.

We first derive the oracle shrinkage coefficient $\rho_2^{\text{OS}}$. Expanding the numerator (hereafter $A$) of Eq. (\ref{eq:risk_minimization}) yields:
\begin{align}
  A
  &= \mathbb{E}(\Trace ( \M C \Fest ))
-\mathbb{E}(\Trace ( \M C \Sest ))
-\mathbb{E}(\Trace ( \Sest\Fest ))
+\mathbb{E}(\Trace ( \Sest^2 ))\nonumber\\
&={\sum}_{i} [\M C]_{ii}^2-\Trace ( \M C^2 )
-{\sum}_{i} \mathbb{E} [\V\Sest]_{ii}^2 + \mathbb{E} ( \Trace ( \Sest^2 ) ) \,.
\end{align}
The variance of the sample variance
$\text{Var} ( [\V\Sest]_{ii}^2 )$ of Gaussian random variables
is equal to $2[\M C]_{ii}^2/N$, and its expectation is
$[\M C]_{ii}$, thus the expectation of the square of the sample
variance $\mathbb{E} ( [\V\Sest]_{ii}^2 )$ is equal to
$2[\M C]_{ii}^2/N + [\M C]_{ii}^2=\frac{N+2}{N}[\M C]_{ii}^2$. The
expectation $\mathbb{E} ( \Trace ( \Sest^2 ) )$ is
given in \cite{chen2010shrinkage}: $\mathbb{E} (\Trace ( \Sest^2
  ) ) =\frac{N+1}{N}\Trace(\M C^2)+\frac{1}{N}\Trace^2(\M C)$.
The numerator then becomes:
\begin{multline}
	A = {\sum}_{i} [\M C]_{ii}^2-\Trace ( \M C^2 )
- N^{-1}(N+2){\sum}_{i} [\M C]_{ii}^2 \hfill\\
+ N^{-1}(N+1)\Trace(\M C^2) + N^{-1}\Trace^2(\M C)\\
~ = \frac{1}{N}\Trace^2(\M C) + \frac{1}{N} \Trace(\M C^2) - \frac{2}{N}{\sum}_{i}
  [\M C]_{ii}^2 \,. ~~~~~~~~~~~~~
\end{multline}
Next, the denominator (hereafter $B$) of Eq. (\ref{eq:risk_minimization}) becomes:
\begin{align}
	B &= 
	   \mathbb{E} ( \Trace(\Sest^2) ) -2\mathbb{E}(\Trace(\Sest \Fest)) + \mathbb{E}(\Trace(\Fest^2))\nonumber\\
&= \frac{N+1}{N}\Trace(\M C^2) + \frac{1}{N}\Trace^2(\M C) - \frac{N+2}{N}\sum_{i} [\M C]_{ii}^2 \,.
\end{align}
Thus, the optimal weight $\rho_2^{\text{OS}}=A/B$ is:
\begin{align}
  \rho_2^{\text{OS}} =
\frac{\Trace(\M C^2)+\Trace^2(\M C)-2\sum_{i}
  [\M C]_{ii}^2}{(N+1)\Trace(\M C^2)+\Trace^2(\M C)-(N+2)\sum_{i}
  [\M C]_{ii}^2}.
\label{eq:rho_oracle_2}
\end{align}


As in Sect. \ref{sec:problem_statement}, the coefficient $\rho_2^{\text{OS}}$ is impractical to evaluate at it depends on the unknown (oracle) covariance $\M C$. Following a similar strategy to \cite{chen2010shrinkage}, we perform a recursive estimation by iteratively plugging a previous estimate of $\M C$ to refine the estimation of $\rho_2^{\text{OS}}$. This recursion over iteration $k$ is given by:
\begin{align}
\widehat\rho_{k+1} &= \displaystyle\frac{\Trace(\widehat{\M C}_k\Sest)+\Trace^2(\widehat{\M C}_k)-2\sum_{i}
  [\widehat{\M C}_k]_{ii}[\Sest]_{ii}}{(N+1)\Trace(\widehat{\M C}_k\Sest)+\Trace^2(\widehat{\M C}_k)-(N+2)\sum\limits_{i}
  [\widehat{\M C}_k]_{ii}[\Sest]_{ii}}\nonumber\\
\widehat{\M C}_{k+1} &= (1-\widehat\rho_{k+1})\Sest + \widehat\rho_{k+1}\Fest\,.
  \label{eq:oas_recursion_2}
\end{align}
Replacing $\widehat{\M C}_{k}$ by its definition and noting that $\Trace(\Fest)=\Trace(\Sest)$ simplify the terms involved in Eq. (\ref{eq:oas_recursion_2}):
\begin{align}
  &\Trace(\widehat{\M C}_k \, \Sest) = (1-\widehat\rho_{k})\Trace(\Sest^2) +
                           \widehat\rho_{k}\Trace(\Fest \, \Sest) \,,\\
  &\Trace^2(\widehat{\M C}_k) = ((1-\widehat\rho_{k})\Trace(\Sest) +
                           \widehat\rho_{k}\Trace(\Fest))^2=\Trace^2(\Sest)\,,\nonumber\\
&[\widehat{\M C}_k]_{ii}[\Sest]_{ii} = [\Sest]_{ii}^2\,\Rightarrow\,{\sum}_i[\Sigest_k]_{ii}[\Sest]_{ii}=\Trace(\Fest\Sest)={\sum}_i[\Sest]_{ii}^2\,.\nonumber
\end{align}
The recursion over the parameter $\widehat\rho_k$ is thus given by:
\begin{equation}
  \widehat\rho_{k+1}({c\, \widehat\rho_{k} +d}) = a\, \widehat\rho_{k} + b\,,
  \label{eq:recursion_polynom}
\end{equation}
where $a=\sum_{i}[\Sest]_{ii}^2-\Trace(\Sest^2)$,
$b=\Trace(\Sest^2)
+\Trace^2(\Sest)-2\sum_{i}[\Sest]_{ii}^2$,
$c=(N+1)\left(\sum_{i}[\Sest]_{ii}^2-\Trace(\Sest^2)\right)$
and
$d=(N+1)\Trace(\Sest^2)-(N+2)\sum_{i}[\Sest]_{ii}^2+\Trace^2(\Sest)$
are 4 constants (independent of $k$).
Recursion (\ref{eq:oas_recursion_2}) converges to $\widehat\rho=1$ or to the fixed-point:
\begin{align}
	\widehat\rho_2^{\text{ OAS}} = \frac{\Trace(\Sest^2)+\Trace^2(\Sest)-2\sum_{i}[\Sest]_{ii}^2}{(N+1)(\Trace(\Sest^2)-\sum_{i}[\Sest]_{ii}^2)} \,,
	\label{eq:rho_oas_2}
\end{align}
which is an extension of the result (\ref{eq:rho_chen}) of \cite{chen2010shrinkage} to the case of a regularization matrix $\widehat{\M F}$ with non-uniform diagonal values.

\section{Shrinkage for non-centered samples\\and weighted statistics}
\label{sec:sample_mean}

\noindent In this section, we further extend the result from Sect. \ref{sec:shrinkage_non_iid}, still focusing on a regularization matrix $\widehat{\M F}$ with non-uniform diagonal values. However, we now consider the general case introduced in Sect. \ref{sec:problem_statement}, where the samples $\{ \V x_n \}_{n=1:N}$ have a non-null (unknown) true mean $\V \mu$. 
 Centering the samples with $\widehat{\V \mu}$ would bias the estimator $\widehat{\rho}_2^{\text{ OAS}}$. 
Additionally, the sample statistics we now consider include weights as introduced in Eq. (\ref{eq:sample_mean_covariances}), and the parameter $\gamma \neq 1$ is included in Eq. (\ref{eq:shrinkage_covariance}) to compensate for the bias introduced by $\widehat{\V \mu}$ in $\widehat{\M S}$. In that general framework, we start by developing the OS estimate given in Eq. (\ref{eq:risk_minimization}):
\begin{equation}
	\rho_3^{\text{OS}} = \frac{\gamma \sum_{i \neq j} \mathbb{E}( [\widehat{\M S}]_{i j}^2) - (1-\epsilon)\sum_{i \neq j} [\M C]_{i j}^2 }{\gamma \sum_{i \neq j} \mathbb{E}([\widehat{\M S}]_{i j}^2)}\,.
	\label{eq:risk_minimization_3}
\end{equation}
Expression (\ref{eq:risk_minimization_3}) requires to compute moments of a Wishart distribution, and we now derive the expression of $\sum_{i \neq j} \mathbb{E}( [\widehat{\M S}]_{i j}^2)$.

When the true mean $\V \mu$ is unknown, the empirical mean $\widehat{\V \mu}$ provides
an estimator. Introducing the centered variables: 
\begin{equation}
  \V z_n = \V x_n - \V \mu ~\text{and}~
  \V \zeta = \widehat{\V \mu} - \V \mu = \big( {\sum}_{n} \alpha_n \V z_n \big) / \big( {\sum}_{n} \alpha_n \big),
\end{equation}
yields:
\begin{equation}
  \V x_n - \widehat{\V \mu} = \V z_n - \V \zeta = \big( {\sum}_{n'} \alpha_{n'}  (\V z_n - \V z_{n'}) \big) / \big( {\sum}_{n'}
  \alpha_{n'} \big)\,.
\end{equation}
Then, noting by $\sum_{\V n} \left( . \right)$ the multiple sums $\sum_{n_1} \sum_{n_2} \sum_{n_3} \sum_{n_4} \sum_{n_5} \sum_{n_6}  \left( . \right)$, we have:
\begin{multline}
	\widehat{\M S}_{i j} = \frac{\sum_{n} \beta_n  \sum_{n'} \alpha_{n'}  \left[\V z_n -
  \V z_{n'}\right]_i  \sum_{n''} \alpha_{n''}  {\left[\V z_n - \V z_{n''}\right]_j} }{ \left(\sum_{n} \beta_n \right) \, \left( \sum_{n} \alpha_n \right)^2}\nonumber\\ \text{so } \mathbb{E}([\widehat{\M S}]_{i j}^2) = C^{-1} {\sum}_{\V n}  \beta_{n_1} \alpha_{n_2} \alpha_{n_3} \alpha_{n_4} \alpha_{n_5} \alpha_{n_6} \, D \,, \hfill
\end{multline}
with $C= \left( \sum_{n} \beta_n \right)^2 \, \left( \sum_{n} \alpha_n \right)^4$ and $D \equiv \V \xi_{i, j, n_1, n_2, n_3, n_4, n_5, n_6}$
\begin{equation}
	D = \mathbb{E} ( [\V z_{n_1} -
   \V z_{n_2}]_i  [ \V z_{n_1} - \V z_{n_3} ]_j  [\V z_{n_4} - \V z_{n_5}]_i 
   [\V z_{n_4} - \V z_{n_6} ]_j )\,. \nonumber
\end{equation}
Developing $D = \V \xi_{i, j, n_1, n_2, n_3, n_4, n_5, n_6}$ yields 16 terms:
\begin{multline}
  \V \xi_{i, j, n_1, n_2, n_3, n_4, n_5, n_6} = \V \psi_{i, j, n_1, n_1, n_4,
  n_4} - \V \psi_{i, j, n_1, n_1, n_4, n_6}\\ - \V \psi_{i, j, n_1, n_1, n_5, n_4} +
  \V \psi_{i, j, n_1, n_1, n_5, n_6} - \V \psi_{i, j, n_1, n_3, n_4, n_4}\\~~ + \V \psi_{i, j, n_1, n_3, n_4, n_6} +
  \V \psi_{i, j, n_1, n_3, n_5, n_4} - \V \psi_{i, j, n_1, n_3, n_5, n_6}\\~~ 
  - \V \psi_{i, j, n_2, n_1, n_4, n_4} + \V \psi_{i, j, n_2, n_1, n_4, n_6} +
 \V \psi_{i, j, n_2, n_1, n_5, n_4}\\~~ - \V \psi_{i, j, n_2, n_1, n_5, n_6}
  + \V \psi_{i, j, n_2, n_3, n_4, n_4} - \V \psi_{i, j, n_2, n_3, n_4, n_6}\\~~ -
  \V \psi_{i, j, n_2, n_3, n_5, n_4} + \V \psi_{i, j, n_2, n_3, n_5, n_6}\,,~~~~~~~~~~~~~~~~
  \label{eq:xi_16}
\end{multline}
\begin{align}
  &\text{with } \V \psi_{i, j, n, n', n'', n'''}
  \equiv \mathbb{E} \left( \V z_{n, i} \V z_{n', j} \V z_{n'',
  i} \V z_{n''', j} \right) \nonumber\\
  &\,\,\,\,= \left\{\begin{array}{ll}
    \left[\M C\right]_{i i} \left[\M C\right]_{j j} + 2 \left[\M C\right]_{i j}^2 & \text{if } n = n' = n'' = n'''\\
    \left[\M C\right]_{i j}^2 & \text{else if }n = n'\text{ and }n'' = n'''\\
    \left[\M C\right]_{i i} \left[\M C\right]_{j j} & \text{else if }n = n''\text{ and }n' = n'''\\
    \left[\M C\right]_{i j}^2 & \text{else if }n = n'''\text{ and }n' = n''\\
    0 & \text{else}
  \end{array}\right.\nonumber \\
  &\,\,\,\,= (\delta_{n, n'} \delta_{n'', n'''} + \delta_{n, n'''} \delta_{n',
  n''}) \, [\M C]^2_{i j} \nonumber \\& ~~~~~~~~~ ~~~~~~~~~  ~~~~~~~~~ ~~~ + \delta_{n, n''} \delta_{n', n'''} \, \left[\M C\right]_{i i} \left[\M C\right]_{j j} \,,
\end{align}
which follows from Isserlis theorem and that holds for any $i$ and $j$ (not just $i \neq j$). 
Hence $\mathbb{E}([\widehat{\M S}]_{i j}^2)$ takes the form:
\begin{equation}
	\mathbb{E}([\widehat{\M S}]_{i j}^2) = \nu \, [\M C]_{i j}^2 + \eta \, [\M C]_{i i} [\M C]_{j j}\,,
	\label{eq:expectation_3_before_sum}
\end{equation}
where $\nu$ and $\eta$ are constants that only depends of the collections of weights
$\lbrace \alpha_n \rbrace_{n=1:N}$ and $\lbrace \beta_n \rbrace_{n=1:N}$, not on $i$ nor on $j$. 
Thus, by injecting expression (\ref{eq:expectation_3_before_sum}) into Eq. (\ref{eq:risk_minimization_3}) yields:
\begin{equation}
	\rho_3^{\text{OS}} = \frac{\sum_{i \neq j} \gamma \nu [\M C]_{i j}^2 + \gamma \eta [\M C]_{i i} [\M C]_{j j} - (1-\epsilon) [\M C]_{i j}^2}{\sum_{i \neq j} \gamma \nu [\M C]_{i j}^2 + \gamma \eta [\M C]_{i i} [\M C]_{j j}}\,.
	\label{eq:rho_oracle_3}	
\end{equation}


As in Sect. \ref{sec:shrinkage_non_iid}, $\rho_3^{\text{OS}}$ is impractical. 
Iteratively substituting a prior estimate of $\M C$ to refine $\widehat{\rho}$ results in the recursion over $k$:

\begin{multline}
		\widehat{\rho}_{k+1} = \\ ~~~\frac{\sum\limits_{i \neq j} \gamma \nu [\widehat{\M C}_k]_{i j}[\widehat{\M S}]_{i j} + \gamma \eta  [\widehat{\M C}_k]_{i i} [\widehat{\M S}]_{j j} - (1-\epsilon) [\widehat{\M C}_k]_{i j}[\widehat{\M S}]_{i j} }{\sum_{i \neq j} \gamma \nu [\widehat{\M C}_k]_{i j}[\widehat{\M S}]_{i j} + \gamma \eta [\widehat{\M C}_k]_{i i} [\widehat{\M S}]_{j j}} 
	\label{eq:oas_recursion_3}
\end{multline}
Replacing $\widehat{\M C}_{k}$ by its definition (\ref{eq:shrinkage_covariance}) leads to:
\begin{align}
	&\sum_{i \neq j} [\widehat{\M C}_k]_{i j}\,[\widehat{\M S}]_{i j} =\sum_{i \neq j} \gamma (1 - \widehat{\rho}_{k}) [\widehat{\M S}]_{i j}^2 + \gamma \widehat{\rho}_{k} [\widehat{\M F}]_{i j} [\widehat{\M S}]_{i j} \label{eq:recursion_3} \\ & = \gamma (1 - \widehat{\rho}_{k}) {\sum}_{i \neq j} [\widehat{\M S}]_{i j}^2 = \gamma (1 - \widehat{\rho}_{k})  \big( \Trace (\Sest^2) - {\sum}_i [ \Sest ]_{ii}^2 \big) \,. \nonumber
\end{align}
The simplification in the second equality comes from the nullity of the off-diagonal terms of $\widehat{\M F}$. Similarly, it comes:
\begin{multline}
	{\sum}_{i \neq j} [\widehat{\M C}_k]_{i i} [\widehat{\M S}]_{j j} = {\sum}_{i \neq j} \gamma (1 - \widehat{\rho}_{k}) [\widehat{\M S}]_{i i} [\widehat{\M S}]_{j j} + \gamma \widehat{\rho}_{k} [\widehat{\M F}]_{i i} [\widehat{\M S}]_{j j} \\
	= \gamma {\sum}_{i \neq j}  [\widehat{\M S}]_{i i} [\widehat{\M S}]_{j j} = \gamma \big( \Trace^2 (\Sest)  - {\sum}_i [ \Sest ]_{ii}^2 \big)\,. 
\end{multline}
The recursion (\ref{eq:oas_recursion_3}) takes a form similar as Eq. (\ref{eq:recursion_polynom}) with: $a = (\gamma \nu + \epsilon -1) (\Trace(\Sest^2)-\sum_{i}[\Sest]_{ii}^2)$, $b = (1-\epsilon -\gamma \nu) (\Trace(\Sest^2)-\sum_{i}[\Sest]_{ii}^2) - \gamma \eta (\Trace^2(\Sest)-\sum_{i}[\Sest]_{ii}^2)$, $c = \gamma \nu (\Trace(\Sest^2)-\sum_{i}[\Sest]_{ii}^2)$, and $d = - \gamma \eta (\Trace^2(\Sest)-\sum_{i}[\Sest]_{ii}^2$. It converges to $\widehat{\rho}_3^{\text{ OAS}} = 1$ or:
\begin{small}
\begin{align}
	\widehat\rho_3^{\text{OAS}} = \frac{(\gamma \nu + \epsilon - 1) (\Trace(\Sest^2)-\sum_{i}[\Sest]_{ii}^2) + \gamma \eta (\Trace^2(\Sest)-\sum_{i}[\Sest]_{ii}^2)}{\gamma \nu (\Trace(\Sest^2)-\sum_{i}[\Sest]_{ii}^2)} \,.
	\label{eq:rho_oas_3}
\end{align}
\end{small}
It remains to derive the expression of $\lbrace \epsilon\,, \gamma\,, \nu\,, \eta \rbrace$ involved in Eq. (\ref{eq:rho_oas_3}). 
The terms of $\mathbb{E}([\widehat{\M S}]_{i j}^2)$ can be computed as follows:
\begin{small}
\begin{multline}
  \frac{\sum_{\V n} \beta_{n_1} \alpha_{n_2} \alpha_{n_3} \beta_{n_4}
  \alpha_{n_5} \alpha_{n_6} \V \psi_{i, j, n_1, n_1, n_4, n_4}}{\left( \sum_{n
  } \beta_n \right)^2  \left( \sum_{n} \alpha_n \right)^4} \\=
  \left( \frac{\sum_{\V n} \beta_{n_1} \alpha_{n_2} \alpha_{n_3} \beta_{n_4}
  \alpha_{n_5} \alpha_{n_6} \delta_{n_1, n_1} \delta_{n_4, n_4}}{\left(
  \sum_{n} \beta_n \right)^2  \left( \sum_{n} \alpha_n \right)^4 }\right.\hfill\\
   \hfill \left. + \frac{\sum_{\V n} \beta_{n_1} \alpha_{n_2} \alpha_{n_3}
  \beta_{n_4} \alpha_{n_5} \alpha_{n_6} \delta_{n_1, n_4}^2}{\left( \sum_{n
  } \beta_n \right)^2  \left( \sum_{n} \alpha_n \right)^4 } \right) [\M C]_{i 
  j}^2\\
	 \hfill+ \frac{\sum_{\V n} \beta_{n_1} \alpha_{n_2} \alpha_{n_3}
  \beta_{n_4} \alpha_{n_5} \alpha_{n_6} \delta_{n_1, n_4}^2}{\left( \sum_{n
  } \beta_n \right)^2  \left( \sum_{n} \alpha_n \right)^4 } [\M C]_{i i}\, 
  [\M C]_{j j}\\
  = [\M C]_{i j}^2 + \frac{\sum_{n} \beta_n^2}{\left( \sum_{n}
  \beta_n \right)^2}  ([\M C]_{i j}^2 + [\M C]_{i i} \cdot [\M C]_{j j})\,,\hfill
  \label{eq:phi_t4_t4}
\end{multline}
\end{small}
The other terms involving the $\V \psi$ quantities of Eq. (\ref{eq:xi_16}) can be computed in a similar fashion as in Eq. (\ref{eq:phi_t4_t4}).
It leads to:
\begin{align}
	\gamma &= (1 - \epsilon)^{-1} =  \bigg( 1 + \frac{\sum_{n } \alpha_n^2}{(\sum_n \alpha_n)^2} - 2
  		\frac{\sum_n \alpha_n \beta_n}{(\sum\limits_n \alpha_n)  (\sum_n \beta_n)} \bigg)\,,\nonumber\\ \vspace{-4mm}
  	\nu &= 1 + 2 \frac{\sum_n \alpha_n^2}{(\sum_n \alpha_n)^2} + \frac{\sum_n
  \beta_n^2}{(\sum_n \beta_n)^2} - 4 \frac{\sum_n \alpha_n \beta_n}{(\sum_n \alpha_n)  \left( \sum_n \beta_n \right)} \nonumber\\ 
  		&\, \hfill +2 \frac{\sum_n \alpha_n^2 \beta_n}{(\sum_n \alpha_n)^2  \left( \sum_n \beta_n \right)} - 4 \frac{\sum_n \alpha_n   \beta_n^2}{(\sum_n \alpha_n)  \left( \sum_n \beta_n \right)^2}\,,\nonumber\\
  		\eta &= \frac{\sum_n \beta_n^2}{(\sum_n \beta_n)^2} + \frac{2 \sum_n
  \alpha_n^2 \beta_n}{(\sum_n \alpha_n)^2  \left( \sum_n \beta_n \right)} - 
  \frac{4 \sum_n \alpha_n \beta_n^2}{(\sum_n \alpha_n)  \left(
  \sum_n \beta_n \right)^2}\nonumber\\ \nonumber &\, \hfill + \frac{(\sum_n \alpha_n^2)^2}{(\sum_n
  \alpha_n)^4} - 4 \frac{(\sum_n \alpha_n^2) (\sum_n \alpha_n 
   \beta_n)}{(\sum_n \alpha_n)^3  \left( \sum_n \beta_n
  \right)}\nonumber\\ &\, \hfill + 2 \frac{\left( \sum_n \alpha_n \beta_n \right)^2 +
  (\sum_n \alpha_n^2) (\sum_n \beta_n^2)}{(\sum_n \alpha_n)^2 
  \left( \sum_n \beta_n \right)^2}\,.
	\label{eq:eta_general_case}
\end{align}
Results (\ref{eq:rho_oas_3}) and (\ref{eq:eta_general_case}) constitute an extension of the results of \cite{chen2010shrinkage} to the generalized estimators introduced in Sect. \ref{sec:introduction}.

\section{Results}
\label{sec:results}

\subsection{Validation on numerical simulations}
\label{subsec:results_simus}

\begin{figure}
	\centering
	\includegraphics[width=0.5\textwidth]{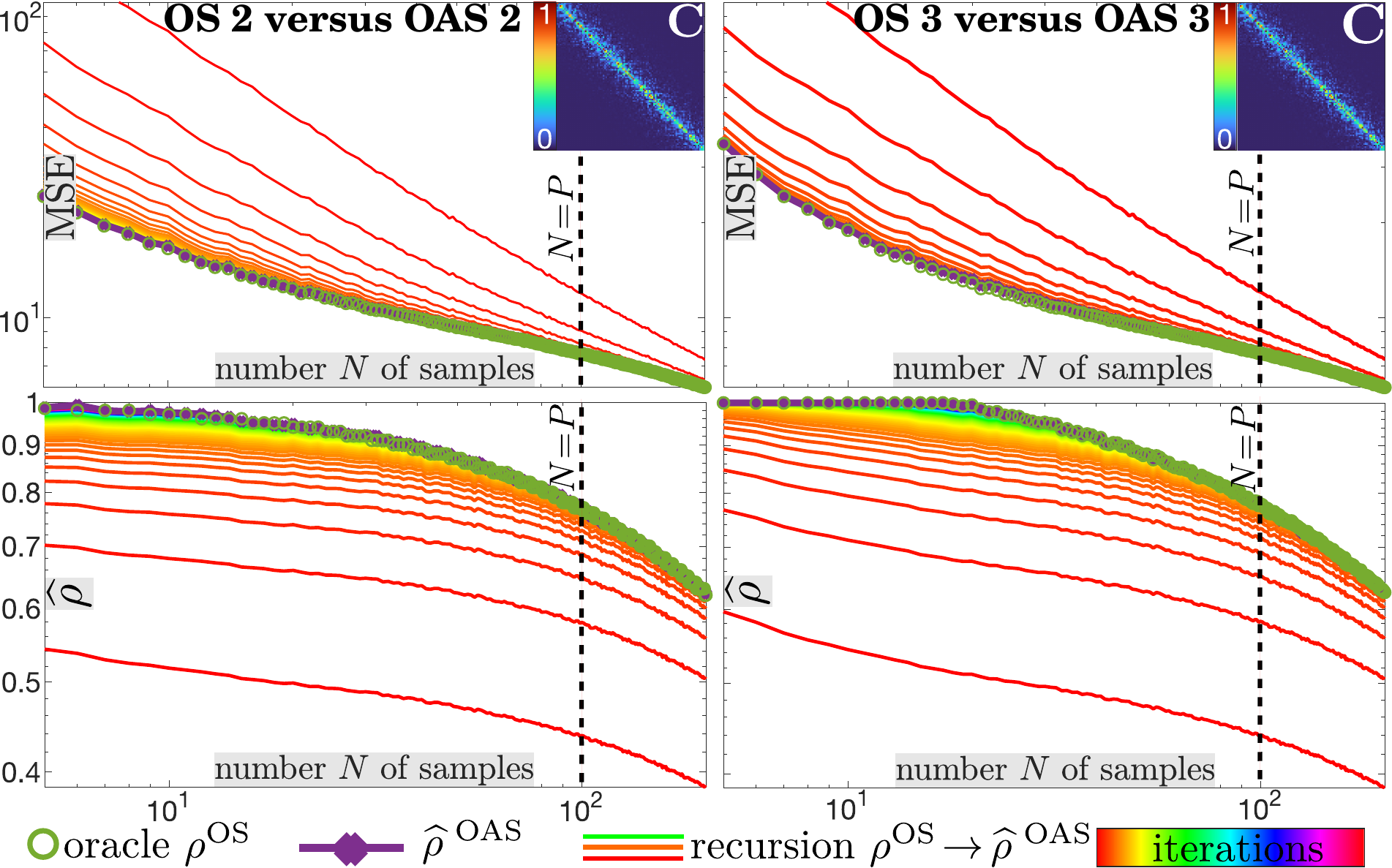}
	\caption{Evolution of the MSE on covariance estimation (top) and of the underlying shrinkage coefficients (bottom) $\widehat{\rho}_2^{\text{ O(A)S}}$ and $\widehat{\rho}_3^{\text{ O(A)S}}$ as a function of $N$ ($P$ fixed at 100). Iterative estimates approximating $\widehat{\rho}_2^{\text{ OS}}$ and $\widehat{\rho}_3^{\text{ OS}}$ by $\widehat{\rho}_2^{\text{ OAS}}$ and $\widehat{\rho}_3^{\text{ OAS}}$ are also reported for comparison.}
	\label{fig:oracle_oas_iterations_fullfig}
\end{figure}

\noindent We aim to validate the derived O(A)S estimators, considering a known covariance matrix $\M C \in \mathbb{R}^{P \times P}$ with $P=100$ (see insert in Fig. \ref{fig:oracle_oas_iterations_fullfig}). Sets of $N$ synthetic multi-variate samples are generated following the targeted statistics $\M C$ via its Cholesky factorization applied to $N$ random i.i.d. Gaussian samples. The resulting empirical covariance $\widehat{\M S}$ is then regularized using shrinkage estimators $\widehat{\rho}_2^{\text{ O(A)S}}$ (Eqs. (\ref{eq:rho_oracle_2})-(\ref{eq:rho_oas_2})) and $\widehat{\rho}_3^{\text{ O(A)S}}$ (Eqs. (\ref{eq:rho_oracle_3})-(\ref{eq:rho_oas_3})). Recursions (\ref{eq:oas_recursion_2})-(\ref{eq:oas_recursion_3}) approximating OS estimates (\ref{eq:rho_oracle_2})-(\ref{eq:rho_oracle_3}) with OAS practical estimators (\ref{eq:rho_oas_2})-(\ref{eq:rho_oas_3}) are also reported. The simulations are repeated $1,000$ times for each tested $N$ value, and we report mean results only as the standard deviation would be smaller than marker size. Figure \ref{fig:oracle_oas_iterations_fullfig} shows the evolution of $\widehat{\rho}$ and resulting mean-square error (MSE) between the shrunk covariance $\widehat{\M C}$ and the true covariance $\M C$ with varying sample size $N$. OAS estimators (\ref{eq:rho_oas_2})-(\ref{eq:rho_oas_3}) asymptotically approximate OS quantities (\ref{eq:rho_oracle_2})-(\ref{eq:rho_oracle_3}), and recursions (\ref{eq:oas_recursion_2})-(\ref{eq:oas_recursion_3}) converge in a few iterations towards OAS estimates. All derived shrinkage coefficients and associated MSE decrease as $N$ increases due to the reduced noise in $\Sest$.

\begin{figure}
	\centering
	\includegraphics[width=0.5\textwidth]{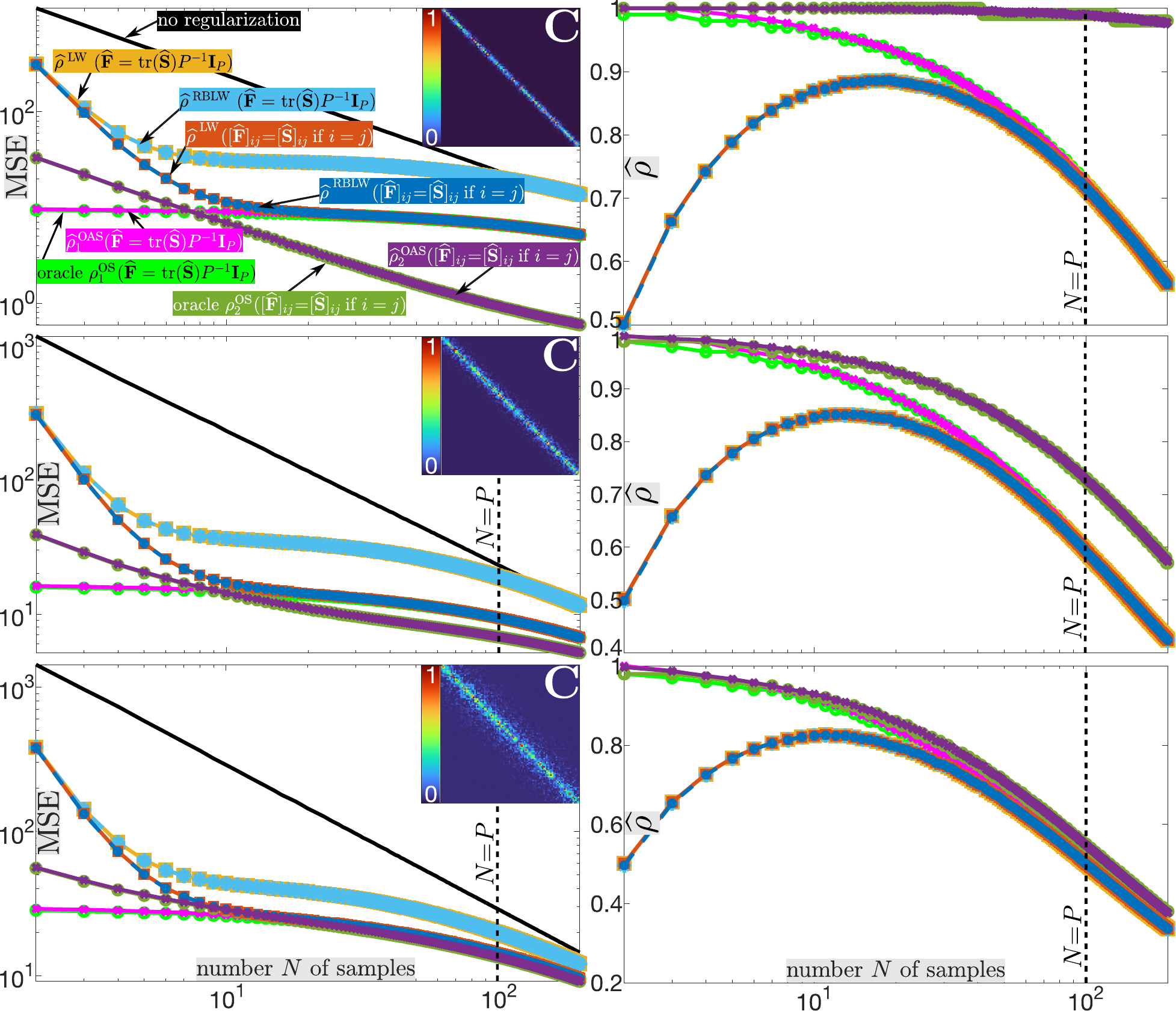}
	\caption{Evolution of the MSE (left) and of shrinkage coefficients (right) 
	$\widehat{\rho}^{\text{ LW}}$, $\widehat{\rho}^{\text{ RBLW}}$, $\widehat{\rho}_1^{\text{ O(A)S}}$, and $\widehat{\rho}_2^{\text{ O(A)S}}$, as a function of $N$ ($P$ fixed at 100). Different true covariance $\M C$ with varying degrees of correlation are considered from top to bottom, see inserts.}
	\label{fig:shrinkage_no_mean_simus_fullfig}
\end{figure}

\begin{figure}[t!]
	\centering
	\includegraphics[width=0.48\textwidth]{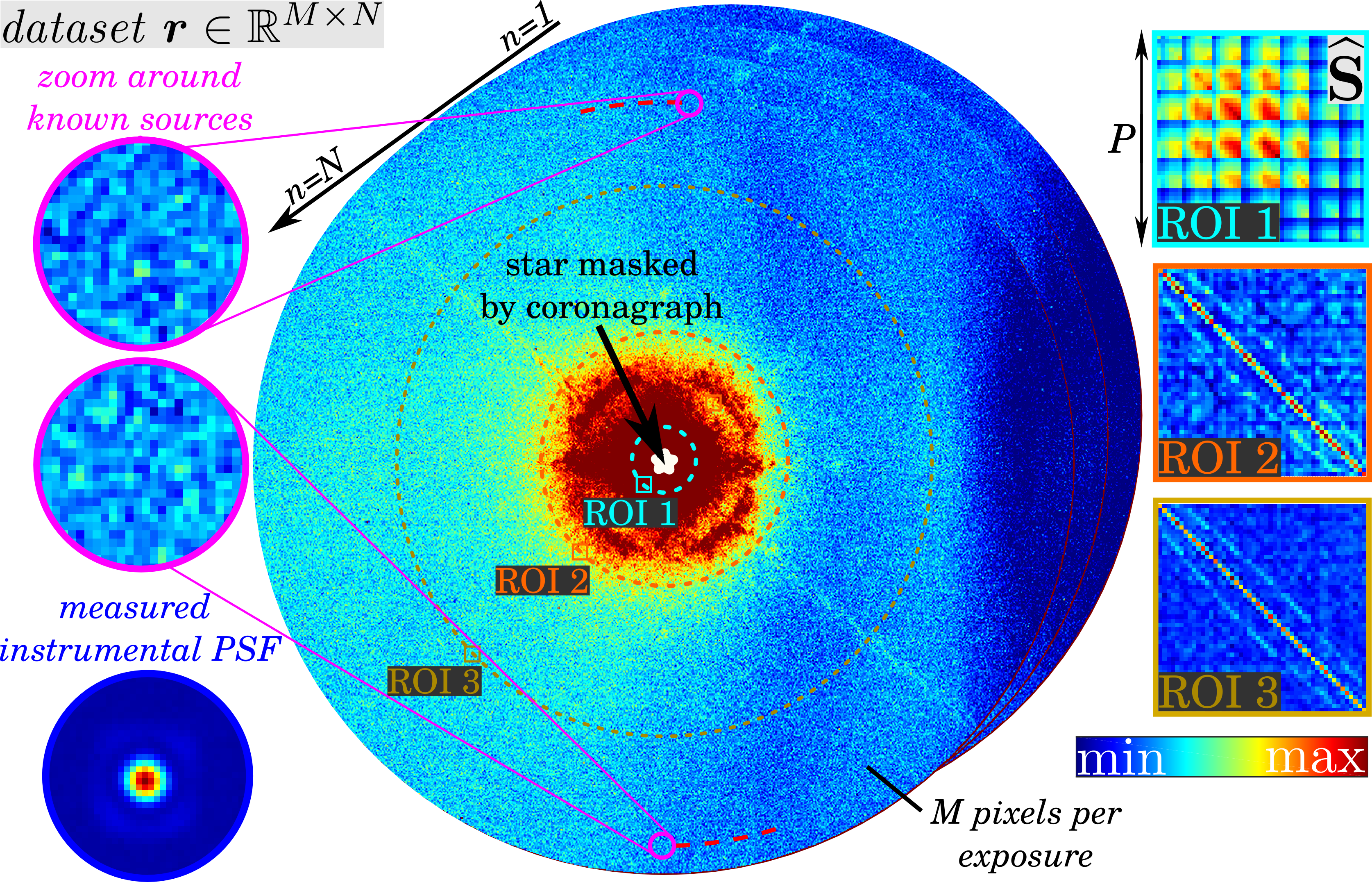}
	\caption{Time-series data at high-contrast. Left top: zooms around two real sources; left bottom: instrument PSF. Right: examples of sample covariance $\Sest$ for different patch ROIs at three distances to the star.}
	\label{fig:data_covariances}
\end{figure}

Now that we have validated our estimators, we compare them to other shrinkage estimators from the literature, assuming centered samples ($\V \mu = \M 0$) and non-weighted empirical statistics ($\alpha_n = \beta_n = 1, \forall n$, and $\gamma = 1$). We thus  focus on $\widehat{\rho}_2^{\text{ O(A)S}}$ estimators in this paragraph (equivalent to $\widehat{\rho}_3^{\text{ O(A)S}}$ under the same working hypothesis). We compare these estimates to the Ledoit-Wolf (LW, \cite{ledoit2004well}) and Rao-Blackwell-Ledoit-Wolf (RBLW, \cite{chen2010shrinkage}) shrinkage methods. 
The LW estimator is a consistent approximation of the oracle (\ref{eq:risk_minimization}) without assumptions on the sample distribution. The RBLW estimator is obtained by computing the conditional expectation of the LW estimator, conditioned on the sufficient statistic $\Sest$, under a Gaussian assumption on the sample distribution. We derive the shrunk covariance from these estimates $\widehat{\rho}^{\text{ (RB)LW}}$ through Eq. (\ref{eq:shrinkage_covariance}). We consider i.i.d. regularization matrices $\widehat{\M F}$ accounting for the average sample variance solely (i.e., $\widehat{\M F} = (\Trace ( \widehat{\M S} ) /P) \M I_P$) as initially proposed by \cite{ledoit2004well}, and also accounting for sample variances (i.e., $[\widehat{\M F}]_{i j} = 0$ if $i \neq j$ and $[\widehat{\M F}]_{i i} = [\widehat{\M S}]_{i i}$ otherwise). Finally, we consider the $\widehat{\rho}_1^{\text{ OAS}}$ estimator (\ref{eq:rho_chen}) from \cite{chen2010shrinkage} as the practical OAS solution of the OS minimization problem (\ref{eq:risk_minimization}) with $\widehat{\M F} = (\Trace ( \widehat{\M S} )/P) \M I_P$. Figure \ref{fig:shrinkage_no_mean_simus_fullfig} presents the evolution of MSE and estimates $\widehat{\rho}$ with varying $N$, following a similar simulation framework as in Fig. \ref{fig:oracle_oas_iterations_fullfig}. Three cases are reported for different true covariances $\M C$ with varying degrees of correlations between sample entries. The RBLW estimator dominates the LW estimator, consistent with trends reported in \cite{chen2010shrinkage}. Accounting for sample variances instead of only average variance in $\widehat{\M F}$ improves MSE for both (RB)LW estimates and our estimators $\widehat{\rho}_2^{\text{ O(A)S}}$ as extensions of $\widehat{\rho}_1^{\text{ O(A)S}}$. Only $\widehat{\rho}_1^{\text{ O(A)S}}$ and $\widehat{\rho}_2^{\text{ O(A)S}}$ are monotonically decreasing functions of $N$; an empirical observation also reported in \cite{chen2010shrinkage} for $\widehat{\rho}_1^{\text{ O(A)S}}$ compared to (RB)LW estimates. Our estimator $\widehat{\rho}_2^{\text{ O(A)S}}$ improves all compared estimators when $N$ is typically larger than 10 (for fixed $P=100$) and seems to converge towards $\widehat{\rho}_1^{\text{ O(A)S}}$ as the number of non-null entries in $\M C$ decreases. In this case, all considered shrinkage estimators converge to the same value of $\widehat{\rho}$ when $N$ is sufficiently large.

\subsection{Application on real data}
\label{subsec:application_astronomy}

\begin{figure*}[t!]
	\includegraphics[width=\textwidth]{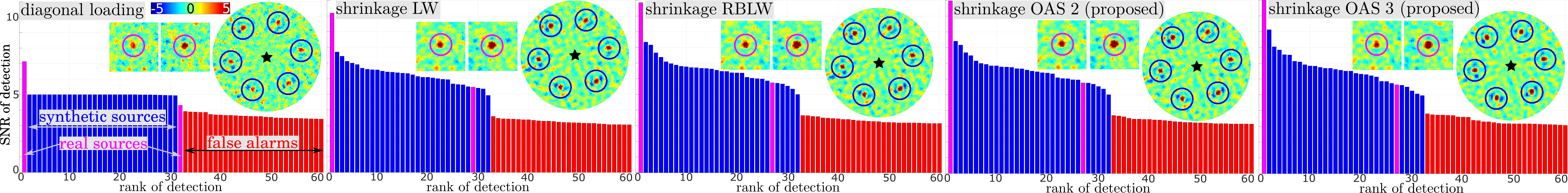}
	\caption{Source detection results: The top 60 detections are visualized as bar charts, arranged by descending SNR values. The two real sources are in pink, 30 synthetic sources in blue, and false detections in red. Insets provide zoomed views of the detection map around the two real sources (pink circles, refer to Fig. \ref{fig:data_covariances} top-left) and the region near the star with six synthetic sources.}
	\label{fig:shrinkage_real_data_fullfig}
\end{figure*}

\noindent We evaluate proposed estimators for a detection problem in astronomy, specifically high-contrast imaging \cite{pueyo2018direct}. This involves observing a star and its nearby region hosting potential exoplanets behaving as point  sources in the data. The observations were carried out with the SPHERE instrument at the Very Large Telescope. It employs adaptive optics to compensate for wavefront distortions induced by atmospheric turbulence, and a coronagraph to block part of the starlight. Nevertheless, faint signals from the sought objects are dominated by a strong and spatially non-stationary \textit{nuisance component} (contrast between the star and exoplanets exceeding $10^{3}$ in images), see Fig. \ref{fig:data_covariances}. This nuisance arises from residual stellar leakages and optical aberrations, resulting in spatially correlated speckles. 
 In a previous work \cite{flasseur2020robustness}, we proposed an exoplanet detection algorithm leveraging the diversity between the quasi-static speckles and the predictable apparent rotation of exoplanets due to the Earth's rotation during observations. The algorithm models non-stationary correlations of the nuisance at the patch scale of $P \simeq 50-100$ pixels. At 2D sky location $\V \theta$, a temporal collection of patches $\V r_{\V \theta} \in \mathbb{R}^{P \times N}$ is modeled by a compound-Gaussian model $\mathcal{N}(\V \mu_{\V \theta}, \{ \alpha^{-1}_{\V \theta, n} \}_{n=1:N} \M C_{\V \theta})$. Parameters $\V \mu_{\V \theta}$ and $\M C_{\V \theta}$ are estimated in a maximum-likelihood sense, along with weights $\{ \alpha_n \}_{n=1:N}$ through a fixed-point iterative scheme. Empirical covariance $\widehat{\M S}$ is regularized by shrinkage. Weights $\alpha_n$ 
 enhance the robustness of mean and covariance estimators against outliers. The impact on shrinkage formulas of their data-driven estimation (possibly from few samples) is left for future work. 
Finally, a matched-filter detects point-like sources, resulting in a signal-to-noise (SNR) detection map. This approach ensures a controlled false alarm probability.

\begin{figure}
	\centering
	\includegraphics[width=0.33\textwidth]{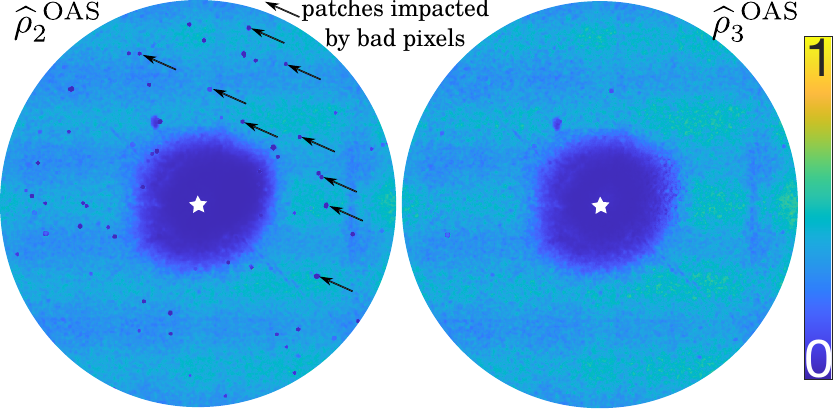}
	\caption{Map of $\widehat{\rho}_2^{\text{ OAS}}$ (left) and $\widehat{\rho}_3^{\text{ OAS}}$ (right) on real data of Fig. \ref{fig:data_covariances}.}
	\label{fig:rho}
\end{figure}

We evaluate the algorithm detection performance for various covariance shrinkage strategies on a real dataset of the HIP 72192 star. The dataset comprises $N = 92$ exposures, each of $M \simeq 10^6$ pixels. To capture most of the PSF energy, circular patches of $P = 113$ pixels are chosen. The resulting $P \times P$ empirical covariances are thus rank deficient. In addition to the two real sources known around this star, we added 30 synthetic faint sources using the direct image formation model. Figure \ref{fig:shrinkage_real_data_fullfig} shows bar diagrams of the SNR of detection for the first 60 highest local maxima in the detection map. Covariance regularization includes diagonal loading (adding $\M \Gamma = 10^{-8} \M I_P$ to $\Sest_{\V \theta}$, whatever $\V \theta$) and spatially adaptive shrinkage with LW, RBLW, OAS 2, and OAS 3 estimators. Zoomed-in detection maps around the real and synthetic sources are provided for each case. SNR-based ranking places the methods as follows: diagonal loading $<$ LW $<$ RBLW $<$ OAS 2 $<$ OAS 3. Examining the two best methods, Fig. \ref{fig:rho} displays maps of the spatially adaptive shrinkage coefficients $\widehat{\rho}_2^{\text{ OAS}}$ and $\widehat{\rho}_3^{\text{ OAS}}$. These coefficients show adaptability to spatial variations. Notably, the $\widehat{\rho}_2^{\text{ OAS}}$ estimator exhibits low shrinkage values in circular patch areas with potential outliers (e.g., bad pixels, numerous on infrared sensors), indicating an impact on shrinkage estimation. This effect is mitigated with the $\widehat{\rho}_3^{\text{ OAS}}$ estimator, which benefits from  $\{ \alpha_n \}_{n=1:N}$ weights to enhance the robustness of mean, covariance, and shrinkage estimators.

\section{Conclusion}
\label{sec:conclusion}


\noindent We introduced generalized shrinkage formulas that can be applied to data with non-stationary variances and non-zero mean. These estimators also incorporate sample weighting to enhance their robustness. We provided practical closed-form expressions for the shrinkage  that asymptotically approximate the oracle. Simulations demonstrate the superiority of these estimators over existing ones, and real data showcases improved sensitivity for exoplanet detection in high-contrast imaging.

\vspace*{2ex}\noindent\begin{minipage}{\columnwidth}
	\footnotesize
\bibliographystyle{IEEEtran}
\bibliography{shrinkage_eusipco_2024_reviewed}
\end{minipage}

\end{document}